\documentstyle[12pt]{article}
\begin{document}
%
%
\begin{titlepage}
\begin{raggedleft}
THES-TP 96/03\\
February 1996\\
\end{raggedleft}
\vspace{2em}
\begin{center}
{\Large\bf{ BRST analysis of the gauged SU(2) WZW \\
\vspace{0.25em}
model and Darboux's transformations}}
\footnote {Work supported by the European Community Human
Mobility program "Eurodafne",Contract CHRX-CT92-00026.}\\
\vspace{2em}
{\large J.E.Paschalis and P.I.Porfyriadis}
\footnote{On leave from the Tbilisi State University, Tbilisi,
Georgia.}
\vspace{1em}\\
Department of Theoretical Physics, University of
Thessaloniki,\\GR 54006 \hspace{0.25em},\hspace{0.25em} Thessaloniki
\hspace{0.25em},\hspace{0.25em} Greece\\
\vspace{1em}
{\scriptsize PASCHALIS@OLYMP.CCF.AUTH.GR\\
PORFYRIADIS@OLYMP.CCF.AUTH.GR}
\end{center}
\vspace{2em}
\begin{abstract}
The four dimensional SU(2) WZW model coupled to elecromagnetism 
is treated as a constraint
system in the context of the BFV approach. We show that the
Darboux's transformations which are used to diagonalize the
canonical one-form in the Faddeev-Jackiw formalism, transform
the fields of the model into BRST invariant ones. The
same analysis is also carried out in the case of spinor
electrodynamics.  
\end{abstract}
\end{titlepage}
%
%
\section{Introduction}
\label{intro}

It is interesting to investigate the relation between the 
Batalin-Fradkin-Vilkovisky (BFV)
quantization \cite{BFV} scheme and the Faddeev-Jackiw approach 
\cite{F-J} to
constrained systems.In the first case the phase space of the theory is
extended by introducing a ghost field for every constraint,
while in the second case the phase space is reduced by
iteratively solving the constraints and performing Darboux's
transformations, until we end up with an unconstraint and
canonical Lagrangian. In the first case the gauge fixing can be
done in an arbitrary way by virtue of the Fradkin-Vilkovisky
theorem while in the second case there is no need for gauge
fixing and we proceed directly to field quantization.

Starting with the BFV formalism one can write the BFV action as
sum of two terms. The first term is the uncanonical one that we
would obtain with the Faddeev-Jackiw method after 
having solved each one of the constraints and the second term is
a BRST exact one. We can see this in
the case of spinor electrodynamics whose Lagrangian density is
given by
\begin{equation}
{\cal L}=-\mbox{\boldmath $\pi$}\bf\cdot\dot{A}\rm
 + i\psi^\dagger\dot{\psi}-H_{0}+
        A_{0}(\nabla\cdot\mbox{\boldmath $\pi$}-\rho)
\end{equation}
where
\begin{displaymath}
H_{0}=\frac{1}{2}(\mbox{\boldmath $\pi$}^2+\bf B^{\rm 2})
-\psi^\dagger\mbox{\boldmath $\alpha$}\cdot(\rm i\nabla+e\bf A \rm)\psi
+m\psi^\dagger\gamma_{0}\psi
\end{displaymath}
See Appendix for notation.

The corresponding BFV action is given by
\begin{equation}
S_{BFV}=\int d^4 x [-\mbox{\boldmath $\pi$}\bf\cdot\dot{A}\rm
 + \pi_{0}\dot{A}_{0}+i\psi^\dagger\dot{\psi}+
\dot{C}{\cal P}+\dot{\bar{C}}\bar{{\cal P}}-H_{0}]+
\int dt [\Psi,Q]
\end{equation}
The scalar potential $A_0$ is promoted to a full dynamical
variable and its conjugate momentum $\pi_0$ has to vanish.
 We have also introduced the canonical
pair $(C,{\cal P})$ of a ghost field and its conjugate momentum,
corresponding to the constraint
$ G_{1}=\rho-\nabla\cdot\mbox{\boldmath $\pi$}$,
and the canonical pair $ (\bar{C},\bar{{\cal P}}) $ of an antighost
field and its canonical momentum, corresponding to the constraint
$ G_{2}=\pi_{0} $. $ \Psi $ is the gauge fermion, and Q is
the BRST charge. The two constraints are first class. The
expression for the BRST charge is given by
\begin{equation}
Q=\int d^3 x [ C(\rho-\nabla\cdot\mbox{\boldmath
$\pi$})+i\bar{{\cal P}}\pi_{0}] 
\end{equation}
and the BRST transformations of the fields are given by
\begin{displaymath}
  s\bf A \rm=-\nabla C \;\; , \;\; \hspace{1em}
      sC=0\;\; , \;\;
\end{displaymath}
\begin{displaymath}
s{\cal P}=\nabla\cdot\mbox{\boldmath $\pi$} -\rho\;\; , \;\;
\hspace{1em} 
      s\mbox{\boldmath $\pi$}=0\;\; , \;\;
\end{displaymath}
\begin{equation}
  sA_{0}=i\bar{{\cal P}}\;\; , \;\; \hspace{1em}
      s\bar{{\cal P}}=0\;\; , \;\;
\end{equation}
\begin{displaymath}
  s\bar{C}=-i\pi_{0}\;\; , \;\; \hspace{1em}
      s\pi_{0}=0\;\; , \;\;
\end{displaymath}
\begin{displaymath}
  s\psi=ieC\psi\;\; , \;\; \hspace{1em}
      s\psi^\dagger=-ieC\psi^\dagger\;\; . \;\;
\end{displaymath}
One can easily see that the canonical Hamiltonian
$ \int d^3 x H_{0} $, and the BFV action $ S_{BFV} $ are BRST
invariant.

We decompose $\bf A $ and $\mbox{\boldmath $\pi$}$ into
transverse and longitudinal components
\begin{displaymath}
 \bf A^{\rm T}=\bf A\rm -\nabla A^{L'}
             \;\; , \;\; \hspace{1em}
     \bf A^{\rm L}\rm=\nabla A^{L'}
            \;\; , \;\; \hspace{1em}
      A^{\rm L'}\rm=\frac{1}{\nabla^2}
(\nabla \cdot \bf A)\rm\;\; , \;\;
\end{displaymath}
\begin{displaymath}
\mbox{\boldmath $\pi$}^{T}=\mbox{\boldmath $\pi$}
-\frac{\nabla}{\nabla^2} \pi^{L'} \;\; , \;\; \hspace{1em}
\mbox{\boldmath $\pi$}^{L}=\frac{\nabla}{\nabla^2} \pi^{L'}\;\;
, \;\; 
\hspace{1em}
      \pi^{L'}=
\nabla \cdot \mbox{\boldmath $\pi$}\rm\;\; . \;\;
\end{displaymath}
Next we use the relations (4) to solve for $C$ , $\bar{{\cal P}}$ ,
 $ \mbox{\boldmath $\pi$}^L$ , $\pi_0$ 
\begin{displaymath}
C=-sA^{L'}\;\; , \;\;
\hspace{1em} 
\bar{{\cal P}}=-isA_{0} \;\; , \;\; \hspace{1em}
\mbox{\boldmath $\pi$}^L =\frac{\nabla}{\nabla^2}(s{\cal P}+\rho)\;\; ,
\;\; \hspace{1em}
\pi_0=is\bar{C}\;\; ,
\end{displaymath}
and we substitute into (2). The resulting expression for
$S_{BFV}$ consists of an uncanonical part and a BRST exact one.
Then we perform the following Darboux's transformations \cite{F-J} 
\begin{displaymath}
\psi \rightarrow \exp(ie A^{L'}) \psi 
\rm \;\; , \;\; \hspace{1em}
  \psi^\dagger \rightarrow \exp(-ie A^{L'}) \psi^\dagger \rm\;\; , \;\;
\end{displaymath}
that diagonalize the uncanonical part. We have then
\begin{eqnarray}  
S_{BFV}&\!\!\!\!\rightarrow&\!\!\!\!\int d^4 x [-\mbox{\boldmath
$\pi$}^T\bf\cdot\dot{A}^{\rm T}\rm
 +i\psi^\dagger\dot{\psi}-H_{C} \nonumber \\
&\!\!\!\!+&\!\!\!\!s[i\bar{C}\dot{A}_{0}+
{\cal P}\dot{A}^{L'}+\frac{1}{2}(s{\cal P})\frac{1}{\nabla^2}{\cal P}+
{\cal P}\frac{1}{\nabla^2}\rho]]+\int dt [\Psi,Q]
\end{eqnarray}
where
\begin{displaymath}
H_{C}=\frac{1}{2}[(\mbox{\boldmath $\pi$}^{\rm T})^2+\bf B^{\rm 2}\rm
-\rho\frac{1}{\bf \nabla \rm^2}\rho]
-\psi^\dagger\mbox{\boldmath $\alpha$}\cdot(\rm i\nabla+e\bf A^{\rm
T} \rm)\psi +m\psi^\dagger\gamma_{0}\psi
\end{displaymath}
is the Coulomb gauge Hamiltonian.
Now we make the following choice for the gauge fermion 
\begin{equation}
\Psi=-\int d^3 x 
[i\bar{C}\dot{A}_{0}+
{\cal P}\dot{A}^{L'}+\frac{1}{2}(s{\cal P})\frac{1}{\nabla^2}{\cal P}+
{\cal P}\frac{1}{\nabla^2}\rho]
\end{equation}
We end up with a canonical unconstrained expression for the
effective action with the longitudinal part of the vector
potential cancelled out.
The same expression for the action would be obtained if we used
the original expression for the Lagrangian density (1) without
extending the phase space of the system, but by merely solving
the constraint $ \nabla\cdot\mbox{\boldmath $\pi$} - \rho =0 $ for 
$\mbox{\boldmath $\pi$}^L $ and rediagonalizing the resulting
expression for the Lagrangian density using the same Darboux's
transformations. This is actually the Faddeev-Jackiw procedure.
It is interesting to note that the Darboux transformed $\psi$
and $\psi^\dagger$ fields are BRST closed and $\sigma$ closed
(physical) where $\sigma$ is the contracting homotopy operator
\cite{Henneaux}. This can
be seen if we perform the previously mentioned Darboux's
transformations in (4). The transverse part of the vector
potential $\bf A^{\rm T}$ which is also present in the
expression for the gauge fixed action is also BRST closed and
$\sigma$ closed. 
%
%
\section{The $U_{EM}(1)$ gauged SU(2) WZW model}
\label{exp-sec}
The $U_{EM}(1)$ gauged 4-dimensional SU(2) WZW model
\cite{Witten,Balachandran} 
 is a phenomenological model which has the symmetries of
electromagnetism and those related to QCD without any extra
ones. It describes the electromagnetic interactions of
pions including those related to the axial anomaly. 

In \cite{PP} this model was treated as a constrained system in the
context of the Faddeev-Jackiw formalism. We expanded the 
effective action into series of powers in the pion fields 
$\theta_{\it a}\;\;,\;\;\it a=\rm 1,2,3$
 and we kept up to second and next
up to third order terms. The Lagrangian density in the first case is
given by
\begin{eqnarray}
  {\cal L}_{eff}&\!\!\!\!=&\!\!\!\!{\cal L}_{EM}+
                 {\cal L}_{\sigma}^{(2)}+{\cal L}_{WZW}^{(2)}
                 +O(\theta^3)\; \; ,\\
  {\cal L}_{EM}&\!\!\!\!=&\!\!\!\!-\frac{1}{4}
                F_{\mu \nu}F^{\mu \nu}\; \; ,\nonumber \\
  {\cal L}_{\sigma}^{(2)}&\!\!\!\!=&\!\!\!\! \frac{1}{2}
               \partial_\mu \theta_{a}\partial^\mu\theta_{a}+
               eA^\mu(\theta_{2}\partial_\mu\theta_{1}-
                      \theta_{1}\partial_\mu\theta_{2})
               +\frac{e^2}{2}A_\mu A^\mu
                (\theta_{1}^{2}+\theta_{2}^{2})\; \; ,\nonumber \\
{\cal L}_{WZW}^{(2)}&\!\!\!\!=&\!\!\!\!-\frac{N_{c}e^2}{12\pi^{2}f_\pi}
               {\epsilon^{\mu \nu \alpha \beta}}A_\mu(\partial_\nu
               A_\alpha)\partial_\beta \theta_{3}\; \; .\nonumber 
\end{eqnarray}

It can be written as an expression first order in time
derivatives as follows
\begin{eqnarray}
{\cal L}_{eff}&\!\!\!\!=&\!\!\!\!-\mbox{\boldmath
$\pi$}\cdot\dot{\bf A}\rm+
                 p_{\it a}\dot{\theta_{\it a}}-H^{(2)}
-A_{0}(\rho^{(2)}-\bf\nabla\cdot \mbox{\boldmath $\pi$}\rm)  
                +O(\theta^3)\; \; ,\\
\vspace{1em}
H^{(2)}_{0} &\!\!\!\!=&\!\!\!\!\frac{1}{2}[\mbox{\boldmath
$\pi$}^2+\bf B^{\rm 2} 
              +(\nabla \rm\theta_{\it a})^{2}+p_{\it a}^{2}]
              +e\bf A\cdot\rm(\theta_{1}\bf \nabla \rm \theta_{2}-
                            \theta_2 \bf \nabla \rm \theta_1)+
               \frac{e^2}{2}\bf A^{\rm 2}\rm
               (\theta_{1}^{2}+\theta_{2}^{2})\nonumber \\
         &\!\!\!\!-&\!\!\!\!\frac{N_{c}e^2}{6\pi^2 f_\pi}\bf
(\mbox{\boldmath $\pi$}\cdot B)\rm\theta_{3}\; \; ,\nonumber \\ 
\vspace{1em}
  \rho^{(2)}&\!\!\!\!=&\!\!\!\! e(p_{2}\theta_1-p_{1}\theta_{2})
                 \; \; ,\nonumber 
\end{eqnarray}
where $p_{\it a}$ are the canonical momenta conjugate to
$\theta_{\it a}$, and $ \rho^{(2)}-\nabla\cdot
\mbox{\boldmath $\pi$}$ 
is the constraint.
We wish to apply the BFV formalism to this model. The BFV action for
the effective Lagrangian density (8) is given by
\begin{equation}
S_{BFV}=\int d^{4} x (-\mbox{\boldmath $\pi$} \cdot \dot{\bf A} \rm +
\pi_{0} \dot{A}_{0} 
+ p_{\it a}\dot{\theta}_{\it a} + \dot{C}{\cal P} + 
\dot{\bar{C}}\bar{{\cal P}}
- H_{0}^{(2)}) + \int d t [\Psi,Q] )
\end{equation}
Here also $\pi_0$ (the conjugate momentum to the scalar
potential $A_0$) has to vanish.
We have introduced, keeping the same notation as in the case of
electrodynamics, the canonical pair 
$(C,{\cal P})$ corresponding to the constraint
$ G_{1}=\rho^{(2)}-\nabla\cdot\mbox{\boldmath $\pi$}$,
and the canonical pair $ (\bar{C},\bar{{\cal P}}) $ corresponding to
the constraint 
$ G_{2}=\pi_{0} $. The two constraints are first class.
The BRST charge is given by 
\begin{equation}
Q=\int d^{3}x [C(\rho^{(2)} - \nabla \cdot \mbox{\boldmath $\pi$}
\rm) +  i \bar{{\cal P}} \pi_{0}]
\end{equation}
and $ S_{BFV} $ is invariant under the BRST transformations
\begin{displaymath}
  s\bf A\rm=-\nabla C\;\; , \;\; \hspace{1em}
      sC=0\;\; , \;\;
\end{displaymath}
\begin{displaymath}
s{\cal P}=\nabla\cdot\mbox{\boldmath $\pi$} -\rho^{(2)}\;\; , \;\;
\hspace{1em} 
      s\mbox{\boldmath $\pi$}=0\;\; , \;\;
\end{displaymath}
\begin{displaymath}
  sA_{0}=i\bar{{\cal P}}\;\; , \;\; \hspace{1em}
      s\bar{{\cal P}}=0\;\; , \;\;
\end{displaymath}
\begin{equation}
  s\bar{C}=-i\pi_{0}\;\; , \;\; \hspace{1em}
      s\pi_{0}=0\;\; , \;\;
\end{equation}
\begin{displaymath}
  s\theta_{1}=-e\theta_{2}C\;\; , \;\; \hspace{1em}
      s\theta_{2}=e\theta_{1}C\;\; , \;\;
\end{displaymath}
\begin{displaymath}
  sp_{1}=-ep_{2}C\;\; , \;\; \hspace{1em}
      sp_{2}=ep_{1}C\;\; , \;\;
\end{displaymath}
\begin{displaymath}
  s\theta_{3}=0\;\; , \;\; \hspace{1em}
      sp_{3}=0\;\; . \;\;
\end{displaymath}
As in the previous case we decompose $\bf A\rm$ and 
$\mbox{\boldmath $\pi$}$ into transverse and longitudinal
components and we solve for C, $\bar{{\cal P}}$ ,
$ \mbox{\boldmath $\pi$}^L $ and $\pi_0$ using relations from (11) 
\begin{displaymath}
C=-sA^{L'}\;\; , \;\;
\hspace{1em} 
\bar{{\cal P}}=-isA_{0} \;\; , \;\; \hspace{1em}
\mbox{\boldmath $\pi$}^L
=\frac{\nabla}{\nabla^2}(s{\cal P}+\rho^{(2)})\;\; , 
\;\; \hspace{1em}
\pi_0=is\bar{C}\;\; .
\end{displaymath}
After substituting in (9) we end up with an expression 
for $ S_{BFV} $ consisting
of an uncanonical part and a BRST exact one as in the case of
electrodynamics. The uncanonical
part is diagonalized by performing the following Darboux's
transformations while the BRST exact one does not change.
\begin{displaymath}
p_1\rightarrow p_1 \cos{\alpha}+p_2\sin{\alpha} \;\; , \;\;
\hspace{1em} 
\theta_1 \rightarrow \theta_1 \cos{\alpha}+\theta_2
\sin{\alpha}\;\; , \;\; 
\end{displaymath}
\begin{displaymath}
p_2 \rightarrow p_2 \cos{\alpha}-p_1\sin{\alpha}\;\; , \;\;
\hspace{1em} 
\theta_2 \rightarrow \theta_2 \cos{\alpha}-\theta_1
\sin{\alpha}\;\; , \;\; 
\end{displaymath}
where $\alpha = eA^{L'}$
 
We have then
\begin{eqnarray}
S_{BFV}&\!\!\!\!\rightarrow&\!\!\!\!\int d^4 x[-\mbox{\boldmath $\pi$}^T
\cdot\dot{\bf A}^{\rm T}+ 
             \rm p_{\it a}\dot{\theta_{\it a}} - H_{C}^{(2)}+sF^{(2)}]
             +\int dt [\Psi,Q]   
     \; \; ,\\
   H_{C}^{(2)}
          &\!\!\!\!=&\!\!\!\!\frac{1}{2}[(\mbox{\boldmath $\pi$}^T)^2
+\bf B^{\rm 2}\rm-
                  \rho^{(2)}\frac{1}{\bf \nabla \rm^2}
                  \rho^{(2)}+(\bf \nabla \rm \rm\theta_{\it a})^{2}
                  +p_{\it a}^{2}]\nonumber \\
          &\!\!\!\!+&\!\!\!\! e\bf A^{\rm T}\cdot\rm
                 (\theta_{1} \bf \nabla \rm \theta_{2}-
                  \theta_2 \bf \nabla \rm \theta_1)
                 + \frac{e^2}{2}(\bf A^{\rm T}\rm)^2
                    (\theta_{1}^{2}+\theta_{2}^{2}) \nonumber \\
          &\!\!\!\!-&\!\!\!\!\frac{N_{c}e^2}{6\pi^2 f_\pi}
                [\mbox{\boldmath $\pi$}^T +
                \frac{\bf \nabla \rm}{\bf \nabla \rm^2}\rm \rho^{(2)}] 
                \cdot\bf B\rm\theta_3 \;\; , \nonumber \\
  F^{(2)} &\!\!\!\!=&\!\!\!\!
  i\bar{C}\dot{A}_{0}+
{\cal P}\dot{A}^{L'}+\frac{1}{2}(s{\cal P})\frac{1}{\nabla^2}{\cal P}+
{\cal P}\frac{1}{\nabla^2}\rho^{(2)}+\frac{N_{c}e^2}{6\pi^2 f_\pi}
(\frac{\nabla}{\nabla^2}{\cal P})\cdot \bf B\rm\theta_3 \nonumber \;\;,
\end{eqnarray}
where $H_{C}^{(2)}$ is the Coulomb gauge Hamiltonian.

Next we take as gauge fermion
\begin{displaymath}
  \Psi=-\int d^3 x F^{(2)}
\end{displaymath}
and we end up with a Coulomb gauge expression for the effective
action with the unphysical $\bf A^{\rm L}$ cancelled out.
In \cite{PP} in the context of the Faddeev-Jackiw formalism we solved
the constraint $ \rho^{(2)}-\nabla\cdot\mbox{\boldmath $\pi$}=0$
 for $\mbox{\boldmath $\pi$}^L$ and
substituted back into the expression (8) for the effective
Lagrangian density. We came up with an uncanonical
expression which was rediagonalized by performing the same
Darboux's transformations. The resulting expression for the
gauge fixed Lagrangian density is the same as the one obtained
in the context of this work.

Replacing the fields by the Darboux transformed ones in (11) we
obtain 
\begin{displaymath}
  sA^{L'}\rm=-C\;\; , \;\; \hspace{1em}
      sC=0\;\; , \;\;
\end{displaymath}
\begin{displaymath}
s{\cal P}=\pi^{L'}-\rho^{(2)}\;\; , \;\;
\hspace{1em} 
      s\pi^{L'}=0\;\; , \;\;
\end{displaymath}
\begin{displaymath}
  sA_{0}=i\bar{{\cal P}}\;\; , \;\; \hspace{1em}
      s\bar{{\cal P}}=0\;\; , \;\;
\end{displaymath}
\begin{equation}
  s\bar{C}=-i\pi_{0}\;\; , \;\; \hspace{1em}
      s\pi_{0}=0\;\; , \;\;
\end{equation}
\begin{displaymath}
  s\theta_{\it a}=0\;\; , \;\; \hspace{1em}
      sp_{\it a}=0\;\; , \;\; \hspace{1em}
\it a=\rm 1,2,3
\end{displaymath}
\begin{displaymath}
  s\bf A^{\rm T}\rm=0\;\; , \;\; \hspace{1em}
      s\mbox{\boldmath $\pi$}^T=0\;\; , \;\;
\end{displaymath}
and
\begin{displaymath}
  \sigma(-C)=A^{L'}\;\; , \;\; \hspace{1em}
      \sigma A^{L'}=0\;\; , \;\;
\end{displaymath}
\begin{displaymath}
\sigma \pi^{L'}={\cal P}\;\; ,
\;\; \hspace{1em} 
      \sigma {\cal P}=0\;\; , \;\;
\end{displaymath}
\begin{displaymath}
  \sigma(i\bar{{\cal P}})=A_{0}\;\; , \;\; \hspace{1em}
      \sigma A_{0}=0\;\; , \;\;
\end{displaymath}
\begin{equation}
  \sigma(-i\pi_{0})=\bar{C}\;\; , \;\; \hspace{1em}
      \sigma(\bar{C})=0\;\; , \;\;
\end{equation}
\begin{displaymath}
  \sigma\theta_{\it a}=0\;\; , \;\; \hspace{1em}
      \sigma p_{\it a}=0\;\; , \;\; \hspace{1em}
\it a =\rm 1,2,3
\end{displaymath}
\begin{displaymath}
  \sigma \bf A^{\rm T}\rm=0\;\; , \;\; \hspace{1em}
      \sigma \mbox{\boldmath $\pi$}^{T}=0\;\; . \;\;
\end{displaymath}
From (13) and (14) we conclude that the Darboux transformed fields
are $\sigma$ and $ s $ closed (physical). The only fields that
are unphysical are actually those that do not appear in the
final expression for the gauge fixed action.

%
%
\section{Keeping third order terms}
\label{exp-third}

Now we proceed with expansion and keep terms up to third order
in $\theta_{\it a}$ \cite{PP}
\begin{equation}
    {\cal L}_{eff}={\cal L}_{EM}+{\cal L}_{\sigma}^{(2)}+
                   {\cal L}_{WZW}^{(2)}+{\cal L}_{WZW}^{(3)}
                   +O(\theta^4)\; \; ,
\end{equation}
where the expression for ${\cal L}_{\sigma}^{(2)}$ and
${\cal L}_{WZW}^{(2)}$ are given in (7) and
\begin{eqnarray}
   {\cal L}_{WZW}^{(3)}&\!\!\!\!=&\!\!\!\!
              -\frac{N_{c}e}{3\pi^{2}f_{\pi}^3}
             {\epsilon^{\mu \nu \alpha \beta}}(\partial_\mu A_\nu)
               (\theta_1\partial_\alpha \theta_2 -
                \theta_2 \partial_\alpha \theta_1)
               \partial_\beta \theta_3\nonumber \\
        &\!\!\!\!+&\!\!\!\! \frac{2N_{c}e^2}{9\pi^{2}f_{\pi}^3}
             {\epsilon^{\mu \nu \alpha \beta}}(\partial_\mu A_\nu)
                      (\partial_\alpha A_\beta)
                (\theta_{1}^2+\theta_{2}^2)\theta_3\nonumber \\
        &\!\!\!\!-&\!\!\!\! \frac{N_{c}e^2}{3\pi^{2}f_{\pi}^3}
                {\epsilon^{\mu \nu \alpha \beta}}
                A_\mu (\partial_\nu A_\alpha)\theta_3
                \partial_\beta (\theta_{1}^2+\theta_{2}^2) 
\nonumber  \; \; .
\end{eqnarray}
We write (15) as an expression first order in time derivatives
\vspace{1em}
\begin{equation}
   {\cal L}_{eff}=-\mbox{\boldmath $\pi$}\cdot \dot{\bf A}\rm +
                  p_{\it a}\dot{\theta_{\it a}}
           - H^{(2)}_{0}-H^{(3)}_{0}-A_{0}(\rho^{(2)}+\rho^{(3)}
           -\nabla\cdot\mbox{\boldmath $\pi$})+O(\theta^4)\; \; ,
\vspace{1em}
\end{equation}
\vspace{1em}
where the expression for $H^{(2)}_{0}$ is given in (8) and
\begin{eqnarray}
\vspace{1em}
   H^{(3)}_{0}&\!\!\!\!=&\!\!\!\!-\frac{N_{c}e}{3\pi^{2}f_{\pi}^3}
              (\mbox{\boldmath $\pi$}\times\bf \nabla 
\rm \theta_3-p_3 \bf B\rm)
            \cdot(\theta_1\bf \nabla \rm \theta_2-\theta_2\bf \nabla
              \rm\theta_1)\nonumber \\
      &\!\!\!\!+&\!\!\!\! \frac{4N_{c}e^2}{9\pi^{2}f_{\pi}^3}
              (\mbox{\boldmath $\pi$}\cdot \bf B\rm)
              ( \theta_{1}^2+\theta_{2}^2)\theta_3 -
              \frac{N_{c}e^2}{3\pi^{2}f_{\pi}^3}
              (\mbox{\boldmath $\pi$}\times\bf A\rm) \cdot [\bf \nabla
              \rm(\theta_{1}^2+\theta_{2}^2)]\theta_3\nonumber \\
      &\!\!\!\!-&\!\!\!\! \frac{2N_{c}e^2}{3\pi^{2}f_{\pi}^3}
              (\bf A\cdot B\rm)(p_1\theta_1+p_2\theta_2)\theta_3
              - \frac{N_{c}e}{3\pi^{2}f_{\pi}^3}
              (\bf B\cdot\bf \nabla \rm\theta_3)
              (p_2\theta_1-p_1\theta_2)\; \; ,\nonumber \\
  \rho^{(2)}&\!\!\!\!=&\!\!\!\! e(p_2\theta_1-p_1\theta_2)
\; \; ,\nonumber \\
  \rho^{(3)}&\!\!\!\!=&\!\!\!\!-\frac{N_{c}e^2}{3\pi^{2}f_{\pi}^3}
                      \bf \nabla \rm\cdot[\bf B\rm
(\theta_{1}^2+\theta_{2}^2)\theta_3] \; \; . \nonumber 
\end{eqnarray}
The BFV effective action is given by 
\begin{equation}
S_{BFV}=\int d^{4} x (-\mbox{\boldmath $\pi$} \cdot \dot{\bf A} \rm +
\pi_{0} \dot{A}_{0} 
+ p_{\it a}\dot{\theta}_{\it a} + \dot{C}{\cal P} + 
\dot{\bar{C}}\bar{{\cal P}}
- H_{0}^{(2)}- H_{0}^{(3)}) + \int dt [\Psi,Q] 
\end{equation}
and
\begin{equation}
Q=\int d^{3}x [C(\rho^{(2)}+\rho^{(3)} - 
\nabla \cdot \mbox{\boldmath $\pi$}
\rm) +  i \bar{{\cal P}} \pi_{0}]
\end{equation}
is the BRST charge.

$S_{BFV}$ is invariant under the BRST transformations
\begin{displaymath}
  s\bf A\rm=-\nabla C\;\; , \;\; \hspace{1em}
      sC=0\;\; , \;\;
\end{displaymath}
\begin{displaymath}
s{\cal P}=\nabla\cdot\mbox{\boldmath $\pi$}
-\rho^{(2)}-\rho^{(3)} \;\; , \;\;
\hspace{1em} 
s\mbox{\boldmath $\pi$}=\frac{N_{c}e^2}{3\pi^{2}f_{\pi}^3}
\nabla[(\theta_{1}^2+\theta_{2}^2)\theta_3]\times\nabla C 
\;\; , \;\;
\end{displaymath}
\begin{displaymath}
  sA_{0}=i\bar{{\cal P}}\;\; , \;\; \hspace{1em}
      s\bar{{\cal P}}=0\;\; , \;\;
\end{displaymath}
\begin{equation}
  s\bar{C}=-i\pi_{0}\;\; , \;\; \hspace{1em}
      s\pi_{0}=0\;\; , \;\;
\end{equation}
\begin{displaymath}
  s\theta_{1}=-e\theta_{2}C\;\; , \;\; \hspace{1em}
 sp_{1}=-ep_{2}C-\frac{2N_{c}e^2}{3\pi^{2}f_{\pi}^3}
(\bf B\rm\cdot\nabla C) \theta_1 \theta_3
      \;\; , \;\;
\end{displaymath}
\begin{displaymath}
  s\theta_{2}=e\theta_{1}C
\;\; , \;\; \hspace{1em}
      sp_{2}=ep_{1}C-\frac{2N_{c}e^2}{3\pi^{2}f_{\pi}^3} 
(\bf B\rm\cdot\nabla C) \theta_2 \theta_3
\;\; , \;\;
\end{displaymath}
\begin{displaymath}
  s\theta_{3}=0\;\; , \;\; \hspace{1em}
      sp_{3}=-\frac{N_{c}e^2}{3\pi^{2}f_{\pi}^3}
(\bf B\rm\cdot\nabla C) (\theta_{1}^2+ \theta_{2}^2)
\;\; . \;\;
\end{displaymath}
We repeate the same procedure as in the previous case by solving 
for C,$\bar{{\cal P}}$,$\mbox{\boldmath $\pi$}^L $ and $\pi_0$
from (19). 
The resulting uncanonical term is diagonalized by the following
Darboux's transformations.
\begin{displaymath}
 \begin{array}{l}
\vspace{0.5em}
\theta_1 \rightarrow \theta_1 \cos{\alpha}+\theta_2 \sin{\alpha}
\;\; , \;\; \\
\vspace{0.5em}
\theta_2 \rightarrow \theta_2 \cos{\alpha}-\theta_1 \sin{\alpha}
\;\; , \;\; \\
\vspace{0.5em}
\mbox{\boldmath $\pi$}^T\rightarrow \mbox{\boldmath $\pi$}^T-
\frac{N_{c}e^2}{3\pi^{2}f_{\pi}^3} 
\bf \nabla \rm[(\theta_{1}^2+\theta_{2}^2)\theta_3]\bf 
                        \times A^{\rm L}\;\;,\\
\vspace{0.5em}
p_1\rightarrow p_1 \cos{\alpha}+p_2\sin{\alpha}
+\frac{2N_{c}e^2}{3\pi^{2}f_{\pi}^3} 
               (\bf B\cdot A^{\rm L}\rm)
(\theta_1 \cos{\alpha}+\theta_2 \sin{\alpha})\theta_3\;\;,\\ 
\vspace{0.5em}
p_2 \rightarrow p_2 \cos{\alpha}-p_1\sin{\alpha}
+\frac{2N_{c}e^2}{3\pi^{2}f_{\pi}^3} 
               (\bf B\cdot A^{\rm L}\rm) 
(\theta_2 \cos{\alpha}-\theta_1 \sin{\alpha})\theta_3\;\;,\\ 
\vspace{0.5em}
  p_3 \rightarrow p_3 +\frac{N_{c}e^2}{3\pi^{2}f_{\pi}^3}
               (\bf B\cdot A^{\rm L}\rm)(\theta_{1}^2+
                 \theta_{2}^2)\;\; . \\
 \end{array}
\end{displaymath}
We have then
\begin{equation}
S_{BFV}\rightarrow\int d^4 x[-\mbox{\boldmath $\pi$}^T
\cdot\dot{\bf A}^{\rm T}+ 
             \rm p_{\it a}\dot{\theta_{\it a}} - H_{C}^{(2)}
-H_{C}^{(3)}+s(F^{(2)}+F^{(3)})]
             +\int dt [\Psi,Q]
\end{equation}
where $ H_{C}^{(2)}$ and $F^{(2)}$ are given in (12),
\begin{eqnarray}
   H_{C}^{(3)}&\!\!\!\!=&\!\!\!\!-\frac{N_{c}e}{3\pi^{2}f_{\pi}^3}
              (\mbox{\boldmath $\pi$}^{T}\times \bf \nabla
 \rm \theta_3-p_3 \bf B\rm)
            \cdot(\theta_1 \bf \nabla \rm \theta_2-\theta_2\bf \nabla
                    \rm\theta_1)\nonumber \\
      &\!\!\!\!+&\!\!\!\! \frac{4N_{c}e^2}{9\pi^{2}f_{\pi}^3}
              (\mbox{\boldmath $\pi$}^{T}\cdot \bf B\rm)
              (\theta_{1}^2+\theta_{2}^2)\theta_3 -
              \frac{N_{c}e^2}{3\pi^{2}f_{\pi}^3}
              (\mbox{\boldmath $\pi$}^{T}\times \bf A^{\rm T}\rm)\cdot
       [\bf \nabla \rm(\theta_{1}^2+\theta_{2}^2)]\theta_3\nonumber \\
      &\!\!\!\!-&\!\!\!\! \frac{2N_{c}e^2}{3\pi^{2}f_{\pi}^3}
              (\bf A^{\rm T}\cdot B\rm)
              (p_1\theta_1+p_2\theta_2)\theta_3 -
              \frac{N_{c}e}{3\pi^{2}f_{\pi}^3}
              (\bf B\cdot\bf \nabla \rm\theta_3)
              (p_2\theta_1-p_1\theta_2)\; \; ,\nonumber 
\end{eqnarray}
and
\begin{eqnarray}
   F^{(3)}&\!\!\!\!=&\!\!\!\!{\cal P}\frac{1}{\nabla^2}\rho^{(3)}
+\frac{N_{c}e}{3\pi^{2}f_{\pi}^3}
   [(\frac{\nabla}{\nabla^2}{\cal P})\times \bf \nabla \rm \theta_3]
           \cdot(\theta_1 \bf \nabla \rm \theta_2-\theta_2\bf \nabla
                    \rm\theta_1)\nonumber \\
      &\!\!\!\!-&\!\!\!\! \frac{4N_{c}e^2}{9\pi^{2}f_{\pi}^3}
              (\frac{\nabla}{\nabla^2}{\cal P})\cdot \bf B\rm
              (\theta_{1}^2+\theta_{2}^2)\theta_3 +
              \frac{N_{c}e^2}{3\pi^{2}f_{\pi}^3}
       [(\frac{\nabla}{\nabla^2}{\cal P})\times \bf A_{\rm T}\rm]\cdot
[\bf \nabla \rm(\theta_{1}^2+\theta_{2}^2)]\theta_3 \;\; . \nonumber  
\end{eqnarray}
Then by fixing the gauge fermion
\begin{displaymath}
  \Psi=-\int d^3 x (F^{(2)}+F^{(3)})
\end{displaymath}
we end up with a Coulomb gauge expression for the effective
action as in the previous case. The same result was obtained in
\cite{PP} using the Faddeev-Jackiw method.

If we perform the Darboux' transformations in (19) we get
\begin{displaymath}
  sA^{L'}\rm=-C\;\; , \;\; \hspace{1em}
      sC=0\;\; , \;\;
\end{displaymath}
\begin{displaymath}
s{\cal P}=\pi^{L'}-\rho^{(2)}-\rho^{(3)}\;\; , \;\;
\hspace{1em} 
      s\pi^{L'}=0\;\; , \;\;
\end{displaymath}
\begin{displaymath}
  sA_{0}=i\bar{{\cal P}}\;\; , \;\; \hspace{1em}
      s\bar{{\cal P}}=0\;\; , \;\;
\end{displaymath}
\begin{equation}
  s\bar{C}=-i\pi_{0}\;\; , \;\; \hspace{1em}
      s\pi_{0}=0\;\; , \;\;
\end{equation}
\begin{displaymath}
  s\theta_{\it a}=0\;\; , \;\; \hspace{1em}
      sp_{\it a}=0\;\; , \;\; \hspace{1em}
\it a=\rm 1,2,3
\end{displaymath}
\begin{displaymath}
  s\bf A^{\rm T}\rm=0\;\; , \;\; \hspace{1em}
      s\mbox{\boldmath $\pi$}^{T}=0\;\; , \;\;
\end{displaymath}
and 
\begin{displaymath}
  \sigma(-C)=A^{L'}\;\; , \;\; \hspace{1em}
      \sigma A^{L'}=0\;\; , \;\;
\end{displaymath}
\begin{displaymath}
\sigma \pi^{L'}={\cal P}\;\; ,
\;\; \hspace{1em} 
      \sigma {\cal P}=0\;\; , \;\;
\end{displaymath}
\begin{displaymath}
  \sigma(i\bar{{\cal P}})=A_{0}\;\; , \;\; \hspace{1em}
      \sigma A_{0}=0\;\; , \;\;
\end{displaymath}
\begin{equation}
  \sigma(-i\pi_{0})=\bar{C}\;\; , \;\; \hspace{1em}
      \sigma(\bar{C})=0\;\; , \;\;
\end{equation}
\begin{displaymath}
  \sigma\theta_{\it a}=0\;\; , \;\; \hspace{1em}
      \sigma p_{\it a}=0\;\; , \;\; \hspace{1em}
\it a=\rm 1,2,3
\end{displaymath}
\begin{displaymath}
  \sigma \bf A^{\rm T}\rm=0\;\; , \;\; \hspace{1em}
      \sigma \mbox{\boldmath $\pi$}^{T}=0\;\; . 
\end{displaymath}
Here again we see that only the fields that appear in the gauge
fixed expression for the effective action are $s$ and $\sigma$
closed (physical).
%
%
\section{Conclusion}
\label{Con}

In this work we drew the analogy between the BFV formalism and
the Faddeev-Jackiw approach in two cases. The spinor
electrodynamics and the 4-dimensional $U_{EM}(1)$ gauged SU(2)
WZW model. According to the BFV formalism the scalar potential
$A_0$ was promoted to a full dynamical variable with vanishing
conjugate momentum $\pi_0$, and the phase space was extended by
introducing a ghost field for every constraint. The BFV action
was written as a sum of an uncanonical term and a BRST exact
one. Darboux's transformations were used to diagonalize the
uncanonical term and the gauge fermion were chosen to cancel the
BRST exact one. The resulting expression for the Coulomb gauge
effective action in both cases are the same as the ones obtaind
by the Faddeev-Jackiw method \cite{PP}.
We also showed that the Darboux transformed fields are BRST and
$\sigma$ closed. 

The SU(3) case is currently under investigation.\\
\vspace{2em}
\\
We wish to thank Dr. Kostas Skenderis for useful discussions.
%
%
\section{Appendix}
\label{app}

Our metric is $g_{\mu \nu}=diag(1,-1,-1,-1) \;\;$.
We choose $ e>0 $. We define $ \epsilon^{0123}=1$ .
 By $\rho$ we
denote the spinor charge density $e\psi^{\dagger} \psi$. 
By $\mbox{\boldmath $\pi$}$ we
denote the electric field $\bf E\rm$ so that
$(\pi_{\mu},A^{\mu}) \;\; \mu=0,1,2,3 $ is a canonical pair. We
made use of the following
Poisson brackets
\begin{eqnarray}
  [A^{\mu}(\bf x,\rm t),\pi^{\nu}(\bf y,\rm t)] &\!\!\!\!=&\!\!\!\! 
  g^{\mu\nu} \delta(\bf x-y \rm) \; \; ,\nonumber \\
\vspace{1em}
  [\theta_{\it a}(\bf x,\rm t),p_{\it b}(\bf y,\rm t)] 
&\!\!\!\!=&\!\!\!\! 
   \delta_{\it ab} \delta(\bf x-y \rm) \; \; ,\nonumber \\
\vspace{1em}
  [\psi_{\it a}(\bf x,\rm t),\psi^{\dagger}_{\it b}(\bf y,\rm t)] 
&\!\!\!\!=&\!\!\!\! 
  -i \delta_{\it ab} \delta(\bf x-y \rm) \; \; ,\nonumber \\
\vspace{1em}
  [C(\bf x,\rm t),{\cal P}(\bf y,\rm t)] &\!\!\!\!=&\!\!\!\! 
  - \delta(\bf x-y \rm) \; \; ,\nonumber \\
\vspace{1em}
[\bar{C}(\bf x,\rm t),\bar{{\cal P}}(\bf y,\rm t)] &\!\!\!\!=&\!\!\!\! 
  - \delta(\bf x-y \rm) \; \; . \nonumber 
\end{eqnarray}
The Grassmann parities of the fields are given by $\;\;\;$
$\epsilon_{A_\mu}=\epsilon_{\pi_\mu}=\epsilon_{\theta_{\it a}}
=\epsilon_{p_{\it a}}=0\;\; ,\;\; 
\epsilon_\psi=\epsilon_{\psi^\dagger}=\epsilon_C=\epsilon_{\cal P}
=\epsilon_{\bar{C}}=\epsilon_{\bar{{\cal P}}}=1$
and their ghost number 
$gh(C)=-gh({\cal P})=1\;\;,\;\; gh(\bar{C})=-gh(\bar{{\cal
P}})=-1 \;\;,\;\;
gh(A_{\mu})=gh(\pi_{\mu})=gh(\theta_{\it a})=gh(p_{\it a})=0$.



\begin{thebibliography}{99}
\bibitem{BFV}
E.S.Fradkin, G.A.Vilkovisky, {\it Phys.Lett.} {\bf B55}
(1975) 224.
I.A.Batalin, G.A.Vilkovisky, {\it Phys.Lett.} {\bf B69}
(1977) 309.
E.S.Fradkin, T.E.Fradkina, {\it Phys.Lett.} {\bf B72}
(1978) 343.
I.A.Batalin, E.S.Fradkin, {\it Phys.Lett.} {\bf B122}
(1983) 157. 
\bibitem{F-J}
L.Faddeev, R.Jackiw, {\it Phys.Rev.Lett.} {\bf 60}
(1988) 1692.\\
 R.Jackiw,{\it (Constraint) Quantization Without Tears, in
DIVERSE TOPICS IN THEORETICAL AND MATHEMATICAL PHYSICS}
( World Scientific, Singapore , 1995 ), hep-th/9306075.
\bibitem{Henneaux}
M.Henneaux, G.Teitelboim, {\it Quantization of Gauge Systems},
p.456 (Princeton University Press, Princeton 1992).
\bibitem{Witten}
E.Witten, {\it Nucl.Phys.} {\bf B223} (1983) 422;\\
 N.K.Pak and P.Rossi, {\it Nucl.Phys.} {\bf B250}
(1985) 279.
\bibitem{Balachandran}
A.P.Balachandran, G.Marmo, B.S.Skagerstam, A.Stern,
{\it Classical Topology and Quantum States}, chapter 16 
(World Scientific, Singapore 1991).
J.F.Donoghue, E.Golowich, B.R.Holstein, {\it Dynamics
of the Standard Model}, p.201. (Cambridge University Press
1992). 
\bibitem{PP}
J.E.Paschalis, P.I.Porfyriadis, {\it Phys.Lett.} {\bf B355}
(1995) 171.

\end{thebibliography}
\end{document}